# Fourier Codes

R. M. Campello de Souza, E. S. V. Freire and H. M. de Oliveira
Department of Electronics and Systems
Federal University of Pernambuco – UFPE, Recife PE, Brazil
eduarda_freire@yahoo.com.br

**Abstract** – A new family of error-correcting codes, called Fourier codes, is introduced. The code parity-check matrix, dimension and an upper bound on its minimum distance are obtained from the eigenstructure of the Fourier number theoretic transform. A decoding technique for such codes is proposed.

## 1. Introduction

THE finite field Fourier transform (FFFT) [1] is a very useful tool in the fields of error control codes and digital signal processing and has been used, for example, as a vehicle to reduce decoder complexity and to implement fast digital convolution [2], [3]. In these scenarios, the existence of fast algorithms for computing the FFFT is a decisive factor for the implementation of such transform domain techniques.

In general, the FFFT is a mapping relating vectors from the Galois field $GF(q)$ to its extension $GF(q^r)$, where $q$ is a prime power $p^m$. When $m = r = 1$, the FFFT is called a Fourier number theoretic transform (FNTT), having applications typically in the field of digital signal processing [4-7].

In this paper, a new family of nonbinary linear block error-correcting codes, called Fourier codes, which are based on the FNTT eigenstructure [8], is introduced. The codewords of a Fourier code are eigensequences of the FNTT. The definition of the eigensequences furnishes the necessary elements to determine the parity check matrix, H, of the code, from which the code parameters $n$ (block length), $k$ (dimension) and a bound on the minimum Hamming distance $d$, are obtained.

This paper is organized as follows. In next section, the eigenstructure of the unitary FNTT is revisited. In Section 3, the Fourier code is introduced and its block length and dimension are obtained. A decoding technique based on the FNTT eigenstructure, to correct single and double errors, is proposed in Section 4. The paper closes with some conclusions on Section 5.

## 2. Eigensequences of the Fourier Number Theoretic Transform

In what follows, the unitary form of the FNTT is considered.

**Definition 1**: The $GF(p)$-valued sequences $x = (x_0, x_1, ..., x_{N-1})$ and $X = (X_0, X_1, ..., X_{N-1})$, form a unitary FNTT pair when

$$X_k = (\sqrt{N})^{-1} (\bmod p) \sum_{n=0}^{N-1} x_n \alpha^{kn}$$

and

$$x_n = (\sqrt{N})^{-1} (\bmod p) \sum_{k=0}^{N-1} X_k \alpha^{-kn},$$

where $\alpha \in GF(p)$ has multiplicative order $N$ and $N^{\frac{p-1}{2}} \equiv 1 (\bmod p)$. The FNTT pair is denoted by $x \leftrightarrow X$ or $X = Fx$, where $F$ is the transform matrix. ∎

A sequence $x$ is said to be an eigensequence of the FNTT, with associated eigenvalue $\lambda \in GF(p^2)$, when it satisfies $X = \lambda x$.

The FNTT and the standard discrete Fourier transform (DFT) have similar eigenstructures. In particular, both satisfy the following lemma [9].

**Lemma 1**: i) The eigenvalues of the FNTT are the fourth roots of unity $(\pm 1, \pm j)$, where $j^2 \equiv -1 (\bmod p)$.
ii) If $x$ is an FNTT eigensequence, then it has even symmetry (i.e. $x_i = x_{N-i}$) if $\lambda \equiv \pm 1 (\bmod p)$ and odd symmetry (i.e. $x_i = -x_{N-i}$) if $\lambda \equiv \pm j (\bmod p)$. ∎

Sequences with even (odd) symmetry are called even (odd) sequences. Such sequences may be used to generate eigensequences according to lemmas 2 and 3 [10].

**Lemma 2**: Let $x \leftrightarrow X$ denote an FNTT pair. Then the sequence $y = E(x) \pm E(X)$ is an eigensequence with eigenvalue $\lambda \equiv \pm 1 (\bmod p)$, where $E(x)$ denotes the even part of $x$. ∎

**Corollary**: Every even sequence $x$ generates an eigensequence $y = x \pm X$. ∎

**Lemma 3**: Let $x \leftrightarrow X$ denote an FNTT pair. Then the sequence $y = O(x) \mp jO(X)$ is an eigensequence with eigenvalue $\lambda = \pm j \pmod{p}$, where $O(x)$ denotes the odd part of $x$. ∎

**Corollary**: Every odd sequence $x$ generates an eigensequence $y = x \mp jX$. ∎

**Example 1**: Consider the GF(5)-valued sequence $x = (4 \; 2 \; 1 \; 4)$ and the 4x4 FNTT matrix over GF(5)

$$F = \begin{pmatrix} 3 & 3 & 3 & 3 \\ 3 & 1 & 2 & 4 \\ 3 & 2 & 3 & 2 \\ 3 & 4 & 2 & 1 \end{pmatrix}.$$

The spectrum of $x$ is $X = (3 \; 2 \; 2 \; 1)$. The even and odd parts of $x$ and $X$ are $E(x) = (4 \; 3 \; 1 \; 3)$, $E(X) = (3 \; 4 \; 2 \; 4)$, $O(x) = (0 \; 4 \; 0 \; 1)$ and $O(X) = (0 \; 3 \; 0 \; 2)$. Therefore $y_1 = E(x) + E(X) = (2 \; 2 \; 3 \; 2)$ and $y_2 = O(x) - 2O(X) = (0 \; 3 \; 0 \; 2)$, respectively, are eigensequences with associated eigenvalues $\lambda \equiv 1 \pmod 5$ and $\lambda \equiv j \equiv 2 \pmod 5$.

## 3. Code Construction

From Definition 1, the $F$ matrix is given by

$$F = (\sqrt{N})^{-1} \pmod p \begin{pmatrix} 1 & 1 & 1 & \cdots & 1 \\ 1 & \alpha & \alpha^2 & \cdots & \alpha^{N-1} \\ 1 & \alpha^2 & \alpha^4 & \cdots & \alpha^{2(N-1)} \\ \vdots & \vdots & \vdots & \ddots & \vdots \\ 1 & \alpha^{N-1} & \alpha^{2(N-1)} & \cdots & \alpha^{(N-1)(N-1)} \end{pmatrix}$$

where $\alpha \in GF(p)$ has order $N$. If $x \leftrightarrow X$ and $x$ is an eigensequence of the linear transform $F$, then its spectrum satisfies $Fx = \lambda x$, so that $(F - \lambda I)x = 0$. As a result, the matrix $(F - \lambda I)$ plays a role analogous to the parity-check matrix of a linear block code with length $N$ and dimension $K$, where $N - K = rank(F - \lambda I)$. In what follows, the standard echelon form of the parity-check and generator matrices is used, i.e., $H = [I_{N-K} | P]$ and $G = [-P^T | I_K]$. Four block codes over $GF(p)$ can be generated, one for each eigenvalue $\lambda$. The possible values for $p$ are determined from the restrictions implied by Definition 1, namely, $N$ is a quadratic residue of $p$ that divides $p - 1$.

**Example 2**: Constructing linear block codes from the FNTT of length $N = 5$, over GF(41). Consider $\alpha = 10$, an element of order 5 in the given field, $\sqrt{5} \equiv 13 \pmod{41}$ and $j \equiv 9 \pmod{41}$. From the transform matrix $F$ one obtains

$$F - \lambda I = \begin{pmatrix} 19-\lambda & 19 & 19 & 19 & 19 \\ 19 & 26-\lambda & 14 & 17 & 6 \\ 19 & 14 & 6-\lambda & 26 & 17 \\ 19 & 17 & 26 & 6-\lambda & 14 \\ 19 & 6 & 17 & 14 & 26-\lambda \end{pmatrix}$$

After some elementary row operations, the parity-check matrices, in standard echelon form, associated with the four eigenvalues $\lambda = \pm 1, \pm j$ are, respectively,

$$H^{(1)} = \begin{pmatrix} 1 & 0 & 0 & 34 & 34 \\ 0 & 1 & 0 & 0 & 40 \\ 0 & 0 & 1 & 40 & 0 \end{pmatrix};$$

$$H^{(-1)} = \begin{pmatrix} 1 & 0 & 0 & 0 & 12 \\ 0 & 1 & 0 & 0 & 40 \\ 0 & 0 & 1 & 0 & 40 \\ 0 & 0 & 0 & 1 & 40 \end{pmatrix};$$

$$H^{(j)} = \begin{pmatrix} 1 & 0 & 0 & 0 & 0 \\ 0 & 1 & 0 & 0 & 1 \\ 0 & 0 & 1 & 0 & 31 \\ 0 & 0 & 0 & 1 & 10 \end{pmatrix};$$

and

$$H^{(-j)} = \begin{pmatrix} 1 & 0 & 0 & 0 & 0 \\ 0 & 1 & 0 & 0 & 1 \\ 0 & 0 & 1 & 0 & 37 \\ 0 & 0 & 0 & 1 & 4 \end{pmatrix},$$

which yields the following Fourier Codes $F(N, K)$ with generator matrices $G^{(\lambda)}$:

$$F(5,2), \quad G_5^{(1)} = \begin{pmatrix} 7 & 0 & 1 & 1 & 0 \\ 7 & 1 & 0 & 0 & 1 \end{pmatrix};$$

$$F(5,1), \quad G_5^{(-1)} = (29 \; 1 \; 1 \; 1 \; 1);$$

$$F(5,1), \quad G_5^{(j)} = (0 \; 40 \; 10 \; 31 \; 1);$$

$$F(5,1), \quad G_5^{(-j)} = (0 \; 40 \; 4 \; 37 \; 1).$$ ∎

## 3.1 The Code Parameters

In a more appropriate notation, consider the Fourier code $F^\lambda(n,k,d)$. The code block length $n$ is the order $N$ of the FNTT matrix. The code dimension $k$ is the multiplicity of the associated eigenvalue $\lambda$, since this is the dimension of the subspace generated by the eigensequences associated with $\lambda$ [9]. The multiplicities of the four eigenvalues are shown in Table 1. Due to the factor $(\sqrt{N})^{-1} \pmod{p}$ in Definition 1, the multiplicity of $\lambda$ depends on the value of $\sqrt{N} \equiv \pm b \pmod{p}$ used. This means that, in Table 1, columns 2 and 3 or 4 and 5 will be interchanged, depending on the value considered ($b$ or $(p-b)$).

**Table 1. Multiplicity of eigenvalues**

| $N$ | Mult. of 1 | Mult. of -1 | Mult. of -j | Mult. of j |
|---|---|---|---|---|
| 4m | m+1 | m | m | m-1 |
| 4m+1 | m+1 | m | m | m |
| 4m+2 | m+1 | m+1 | m | m |
| 4m+3 | m+1 | m+1 | m+1 | m |

It can also be observed that Fourier codes, asymptotically, have rates equal to ¼.

A bound on the minimum distance $d$ is demonstrated in Proposition 1.

**Proposition 1**: Let $H^{(\lambda)} = [I_{n-k} | P]$ be the parity-check matrix of an $F^\lambda(n,k,d)$ Fourier code over $GF(p)$. Then the submatrix $P$ contains a secondary diagonal matrix $D_s$, of order k, with entries $m$, where $m = \begin{cases} p-1, & \text{if } \lambda \equiv \pm 1 \pmod{p}, \\ 1, & \text{if } \lambda \equiv \pm j \pmod{p}. \end{cases}$

**Proof**: According to Lemma 1(ii), $x$ is an eigensequence with even symmetry if its associated eigenvalue is $\lambda \equiv \pm 1 \pmod{p}$. and with odd symmetry if $\lambda \equiv \pm j \pmod{p}$. Thus, if $x \in F^\lambda(n,k,d)$, we have:
i) For $\lambda \equiv \pm 1 \pmod{p}$;
In this case, the codeword $x$ may be written as

$$x = (x_0, x_1, x_2, x_3, \ldots, x_{N-3}, x_{N-2}, x_{N-1}) =$$
$$= (x_0, x_1, x_2, x_3, \ldots, x_3, x_2, x_1).$$

In systematic form, $x = (c_1, c_2, \ldots, c_{n-k}, k_1, k_2, \ldots k_k)$, where $c_i$ and $k_j$ represent the parity-check and information symbols, respectively. This leads to the parity check equations $c_2 = k_k$, $c_3 = k_{k-1}$, $c_4 = k_{k-2}$, ...., $c_{k-1} = k_3$, $c_k = k_2$, and $c_{k+1} = k_1$. These equations, which do not represent the complete set of parity check equations of the code, in matrix form, correspond to the matrix $D_s$.
ii) For $\lambda \equiv \pm j \pmod{p}$;
In a similar way, the condition

$$x = (x_0, x_1, x_2, x_3, \ldots, x_{N-3}, x_{N-2}, x_{N-1}) =$$
$$= (x_0, x_1, x_2, x_3, \ldots, -x_3, -x_2, -x_1),$$

leads to the parity-check equations $c_2 = -k_k$, $c_3 = -k_{k-1}$, $c_4 = -k_{k-2}$, ....., $c_{k-1} = -k_3$, $c_k = -k_2$, $c_{k+1} = -k_1$ and the result follows. ∎

**Example 3**: i) Consider the code $F^1(8,3,4)$ over GF(17), described by the parity-check matrix

$$H^{(1)} = \begin{bmatrix} 1 & 0 & 0 & 0 & 0 & 3 & 5 & 3 \\ 0 & 1 & 0 & 0 & 0 & 0 & 0 & 16 \\ 0 & 0 & 1 & 0 & 0 & 0 & 16 & 0 \\ 0 & 0 & 0 & 1 & 0 & 16 & 0 & 0 \\ 0 & 0 & 0 & 0 & 1 & 14 & 5 & 14 \end{bmatrix}$$

Since $\lambda=1$, $x$ is an eigensequence with even symmetry, $x = (x_0, x_1, x_2, x_3, x_4, x_3, x_2, x_1) = (c_1, c_2, c_3, c_4, c_5, k_1, k_2, k_3)$. Considering $xH^T = 0$, the following parity-check equations arise:

$c_1 + 3k_1 + 5k_2 + 3k_3 = 0$;
$c_2 + 16k_3 = 0$;
$c_3 + 16k_2 = 0$;
$c_4 + 16k_1 = 0$; and
$c_5 + 14k_1 + 5k_2 + 14k_3 = 0$,

which leads to the indicated $D_s$ matrix.

ii) Consider the code $F^j(8,2,4)$ over GF(17), with $j = 4$, described by the parity-check matrix

$$H^{(j)} = \begin{bmatrix} 1 & 0 & 0 & 0 & 0 & 0 & 0 & 0 \\ 0 & 1 & 0 & 0 & 0 & 0 & 0 & 1 \\ 0 & 0 & 1 & 0 & 0 & 0 & 1 & 0 \\ 0 & 0 & 0 & 1 & 0 & 0 & 6 & 1 \\ 0 & 0 & 0 & 0 & 1 & 0 & 0 & 0 \\ 0 & 0 & 0 & 0 & 0 & 1 & 11 & 16 \end{bmatrix}$$

In this case $x = (x_0, x_1, x_2, x_3, x_4, -x_3, -x_2, -x_1) = (c_1, c_2, c_3, c_4, c_5, c_6, k_1, k_2)$ and the following parity check equations arise:

$c_1 = 0$;
$c_2 + k_2 = 0$;
$c_3 + k_1 = 0$;
$c_4 + 6k_1 + k_2 = 0$;
$c_5 = 0$; and
$c_6 + 11k_1 + 16k_2 = 0$,

which leads to the indicated $D_s$ matrix. ∎

**Proposition 2** (An upper bound on the minimum distance): The minimum distance of a Fourier code $F^\lambda(n,k,d)$ satisfies $d \leq n-2k+2$.

**Proof**: According to Proposition 1, there are at least $k-1$ zeros in every column of the $P$ submatrix. Therefore, the maximum weight of these columns is $w_{\max} = (n-k)-(k-1) = n-2k+1$. Considering the generator matrix $G = \left[-P_\lambda^T \mid I_k\right]$, we have $d \leq n-2k+2$. ∎

**Corollary:** In $F^\lambda(n,k,d)$, with $\lambda = \pm j$, the code minimum distance satisfies $d \leq n-2k$. Table 2 shows the values of $n$, $k$ and $d$ for a few Fourier codes.

**Table 2. Parameters of some Fourier codes** $(N, k^\lambda, d^\lambda)$ **for** $\lambda \equiv \pm 1, \pm j \pmod p$.

| N  | $k^{+1}$ | $d^{+1}$ | $k^{-1}$ | $d^{-1}$ | $k^{+j}$ | $d^{+j}$ | $k^{-j}$ | $d^{-j}$ |
|----|----------|----------|----------|----------|----------|----------|----------|----------|
| 3  | 1        | 3        | 1        | 3        | -        | -        | 1        | 2        |
| 4  | 2        | 2        | 1        | 4        | -        | -        | 1        | 2        |
| 5  | 2        | 3        | 1        | 5        | 1        | 4        | 1        | 4        |
| 6  | 2        | 4        | 2        | 4        | 1        | 4        | 1        | 4        |
| 7  | 2        | 5        | 2        | 5        | 1        | 6        | 2        | 4        |
| 8  | 3        | 4        | 2        | 4        | 1        | 6        | 2        | 4        |
| 9  | 3        | 3        | 2        | 6        | 2        | 6        | 2        | 6        |
| 10 | 3        | 6        | 3        | 6        | 2        | 6        | 2        | 6        |
| 11 | 3        | 7        | 3        | 7        | 2        | 8        | 3        | 6        |
| 12 | 4        | 4        | 3        | 6        | 3        | 4        | 2        | 6        |

# 4. Error Control based on the FNTT Eigenstructure

Cyclic codes are linear block codes and can be decoded by any standard decoding technique used for such codes. However, the additional mathematical structure that they have allows the construction of new and more efficient decoding algorithms. Here the same strategy is pursued and the search is for decoding methods based on the FNTT eigenstructure. In this scenario, the symmetry conditions and the possibility of using fast algorithms for computing the FNTT plays an important role.

To illustrate the basic ideas, two distinct situations are considered, single and double error correction. In any case, the decoding steps depend whether the received sequence is symmetric or not.

## 4.1 Syndrome computation

Consider the received sequence to be $r = x + e$, where $x \in F^\lambda(n,k,d)$ and $e$ is the error sequence. The syndrome of $r$ is defined as $S = Fr - \lambda r$, which is zero if, and only if, $r$ is an FNTT eigensequence associated with the eigenvalue $\lambda$. A fast transform can be used to compute $S$.

## 4.2 Single Error Correction

The decoding algorithm is described considering the possible symmetry of the received sequence.

### 4.2.1 The symmetric case (even or odd symmetry)

Consider that the received sequence $r = (r_0, r_1, r_2,..., r_{N-1})$ exhibits the same type of symmetry displayed by the eigensequences associated with the eigenvalue $\lambda$. In what follows, without loss of generality, $\lambda = \pm 1$ is considered. The procedure for $\lambda = \pm j$ follows similar lines.

If $N$ is odd and a single error has occurred, then it must be in the first position of $r$, since only an even number of errors could preserve symmetry. This condition is checked simply by observing that, according to Definition 1, $r_0$ must satisfy

$$r_0 = (\lambda\sqrt{N} - 1)^{-1}(r_1 + r_2 + ... + r_{N-1}),$$

for $r$ to be an eigensequence. If $N$ is even, for the same reason, the possible single-error occurred at position $r_0$ or $r_{N/2}$. Therefore, to decode the received sequence, use the following algorithm:

1. Make $r_0 = (\lambda\sqrt{N} - 1)^{-1}(r_1 + r_2 + ... + r_{N-1})$;
2. If the obtained sequence is an eigensequence, decoding is complete. Otherwise, make $r_{N/2} = r_0(\lambda\sqrt{N} - 1) - (r_1 + ... + r_{N/2-1} + r_{N/2+1} + ... + r_{N-1})$;
3. If the obtained sequence is an eigensequence, decoding is complete. Otherwise, more than one error has occurred.

### 4.2.2 The nonsymmetric case

Consider the received sequence $r = (r_0, r_1, r_2, ..., r_{N-2}, r_{N-1})$. If the symbols $r_i$ and $r_{N-i}$ are different, the error occurred either at position $r_i$ or position $r_{N-i}$. The decoding algorithm is:
1. Substitute $r_i$ by $r_{N-i}$;
2. If the obtained sequence is an eigensequence, decoding is complete. Otherwise, substitute $r_{N-i}$ by $r_i$.
3. If the obtained sequence is an eigensequence, decoding is complete. Otherwise, more than one error has occurred.

## 4.3 Double Error Correction

In what follows, the occurrence of a single error is considered to be a more likely event than the occurrence of double errors. Therefore, for a double error correcting code, the decoding algorithm consists in applying the single error correcting approach described in last section. If decoding is not possible, the occurrence of a double error is assumed and decoding proceeds according to the following steps. Without loss of generality, $N$ is considered to be odd. When $N$ is even, the procedure follows similar lines. In particular, once symmetry is reestablished, the symbol $r_{N/2}$ can be computed from the symmetric symbols via its parity-check equation.

### 4.3.1 The symmetric case

If symmetry is preserved in the received word, the errors have occurred in a pair of symbols ($r_i$, $r_{N-i}$), $i \neq 0$. The decoding algorithm is:

1. For $1 \leq i \leq \frac{N-1}{2}$ make

$$r_i = \frac{1}{2}(\lambda\sqrt{N}r_0 - \sum_{\substack{j=0 \\ j \neq i, N-i}}^{N-1} r_j)$$

and $r_{N-i} = r_i$.

2. If the obtained sequence is an eigensequence, decoding is complete. Otherwise, more than two errors have occurred.

### 4.3.2 The nonsymmetric case

Consider the received sequence $r = (r_0, r_1, r_2, ....r_{N-2}, r_{N-1})$. If a double error has occurred, the possible options are:

i) Errors in symbols $r_0$ and $r_i$, $i \neq 0$.

Decoding Algorithm 1:
1. Substitute $r_i$ by $r_{N-i}$.
2. Make $r_0 = (\lambda\sqrt{N} - 1)^{-1}(r_1 + r_2 + ... + r_{N-1})$;
3. If the obtained sequence is an eigensequence, decoding is complete. Otherwise substitute $r_{N-i}$ by $r_i$ and make $r_0 = (\lambda\sqrt{N} - 1)^{-1}(r_1 + r_2 + ... + r_{N-1})$;
4. If the obtained sequence is an eigensequence, decoding is complete. Otherwise use decoding algorithm 2.

**Example 4**: Consider the double-error correcting code $F^I(7,2,5)$ over GF(29), with $\alpha = 7$, $\sqrt{7} \equiv 6 \pmod{29}$, generator matrix

$$G_7^{(1)} = \begin{pmatrix} 16 & 0 & 1 & 10 & 10 & 1 & 0 \\ 20 & 1 & 0 & 20 & 20 & 0 & 1 \end{pmatrix},$$

and received sequence $r = (16, 2, 1, 10, 10, 1, 3)$.
Substituting $r_1$ by $r_{N-1}$ and making $r_0 = (1.6 - 1)^{-1}(3 + 1 + 10 + 10 + 1 + 3) \equiv 23 \pmod{29}$
results in $r^{(1)} = (23, 3, 1, 10, 10, 1, 3)$ and $R^{(1)} = (23, 22, 25, 25, 25, 25, 22)$.
Substituting $r_{N-1}$ by $r_1$ and making $r_0 = (1.6 - 1)^{-1}(2 + 1 + 10 + 10 + 1 + 2) \equiv 11 \pmod{29}$
results in $r^{(2)} = (11, 2, 1, 10, 10, 1, 2)$ and $R^{(2)} = (11, 5, 17, 20, 20, 17, 5)$. In this case the sequence could not be decoded using Algorithm 1. ∎

ii) One error in $r_i$ and another in $r_{N-i}$, $i \neq 0$. The decoding algorithm is similar to the symmetric case, with the difference that now the error positions are known

Decoding Algorithm 2:
1. Make $r_i = \frac{1}{2}(\lambda\sqrt{N}r_0 - \sum_{\substack{j=0 \\ j \neq i, N-i}}^{N-1} r_j)$ and $r_{N-i} = r_i$.

2. If the obtained sequence is an eigensequence, decoding is complete. Otherwise more than two errors have occurred.

**Example 5**: Considering the decoding of the sequence given in Example 4, we have

$$r_{N-1} = r_1 = \frac{1}{2}(1.6.16 - (16 + 1 + 10 + 10 + 1)) = 0.$$

Therefore $r^{(1)} = (16, 0, 1, 10, 10, 1, 0)$, $R^{(1)} = (16, 0, 1, 10, 10, 1, 0)$ and decoding is complete. ∎

iii) One error in $r_i$, $i \neq 0$, and another in $r_j$, $j \neq N - i$. In this case it is necessary to make at least $2^2$ substitutions to satisfy the symmetry and verify the eigensequence condition.

Decoding Algorithm 3:
1. Substitute $r_i$ by $r_{N-i;}$ and substitute $r_j$ by $r_{N-j;}$
2. If the obtained sequence is an eigensequence, decoding is complete. Otherwise substitute $r_{N-i}$ by $r_i$ and $r_j$ by $r_{N-j;}$
3. If the obtained sequence is an eigensequence, decoding is complete. Otherwise substitute $r_i$ by $r_{N-i}$ and $r_{N-j}$ by $r_{j;}$
4. If the obtained sequence is an eigensequence, decoding is complete. Otherwise substitute $r_{N-i}$ by $r_i$ and $r_{N-j}$ by $r_{j;}$
5. If the obtained sequence is an eigensequence, decoding is complete. Otherwise more than two errors have occurred.

**Example 6**: In Example 4, let the received sequence be $r = (16, 2, 3, 10, 10, 1, 0)$. It is straightforward to verify that at least two errors have occurred. Thus, using the latter algorithm, we have:
First possibility: the decoded sequence is $r^{(1)} = (16, 2, 3, 10, 10, 3, 2)$ and its spectrum $R^{(1)} = (27, 13, 19, 17, 17, 19, 13)$;
Second possibility: the decoded sequence is $r^{(2)} = (16, 2, 1, 10, 10, 1, 2)$ and its spectrum $R^{(2)} = (7, 1, 13, 16, 16, 13, 1)$;
Third possibility: the decoded sequence is $r^{(3)} = (16, 0, 3, 10, 10, 3, 0)$ and its spectrum $R^{(3)} = (7, 12, 7, 11, 11, 7, 12)$;

Fourth possibility: the decoded sequence is $r^{(4)} =$ (*16, 0, 1, 10, 10, 1, 0*) and its spectrum $R^{(4)} =$ (*16, 0, 1, 10, 10, 1, 0*). So, $r_4$ is the estimated transmitted sequence. ∎

In the decoding procedures just described, the algorithms requires only FNTT computations, which can be implemented via fast Fourier transforms, and simple operations such as additions, multiplication by constants and substitutions. Considering that the FNTT of a symmetric sequence displays the same type of symmetry, it is also possible to use Goertzel like algorithms [12], instead of an FNTT computation, to verify the eigensequence condition.

## 5. Conclusions

In this paper a new family of nonbinary linear block codes, the Fourier codes, was introduced. The codewords of a Fourier code $F^\lambda(n,k,d)$ are the eigensequences of the Fourier number theoretic transform, associated with a given eigenvalue $\lambda$. Strategies for single and double error control based upon the FNTT eigenstructure were examined. The approach described in the paper can be extended to other families of finite field transforms, such as the Hartley number theoretic transform [11]. The family of Hartley codes over $GF(p)$ can be constructed following the same approach described here. The families of Fourier and Hartley codes are members of a new class of codes, which we call transform codes. For a given finite field transform of length $N$, its eigenstructure can be used to construct a linear code of length $N$ and dimension $k$, where $k$ is the multiplicity of the transform eigenvalues. Different multiplicities will lead to codes with different rates. In this scenario fast transforms and properties of the transform eigenstructure can assist in the implementation of the code.

Using the same approach and discrete transforms defined over infinite fields, codes defined over the field of real numbers can also be constructed.

The restrictions on the code parameters that results from the use of standard transforms, either finite or infinite, can be removed if arbitrary linear transforms are considered. In this case, codes with unrestrained rates can be constructed.